\documentclass[notitlepage, twocolumn, prl, nobalancelastpage, amssymb, superscriptaddress, showpacs, preprintnumbers, nofootinbib]{revtex4-2}
\usepackage[utf8]{inputenc}
\usepackage[normalem]{ulem}
\usepackage[dvipsnames]{xcolor}
\usepackage{cancel}

\definecolor{red}{rgb}{0.9, 0,0}
\definecolor{cerulean}{rgb}{0., 0.42,0.9}
\definecolor{navy}{rgb}{0.05, 0.05,0.8}

\usepackage[colorlinks]{hyperref}
\hypersetup{
colorlinks = true,
linkcolor = red,
urlcolor  = blue,
citecolor = blue,
anchorcolor = blue
}

\usepackage{amsmath,amsfonts,amsthm,amssymb, bm, mathtools, latexsym} 
\usepackage{hyperref}
\usepackage{subcaption}
\usepackage{physics}
\usepackage[dvipsnames]{xcolor}
\usepackage{enumerate}
\usepackage{slashed}
\usepackage[final]{showlabels}

\usepackage{tikz}
\usepackage{tikz-3dplot}
\usepackage{tikz-feynman,contour}
\usetikzlibrary {positioning}

\def\e{{\epsilon}}

\def\ka{{\kappa}}
\def\o{{\omega}}
\def\a{{\alpha}}
\def\b{{\beta}}
\def\g{{\gamma}}
\def\l{{\lambda}}

\def\d{{\delta}}
\def\t{{\theta}}
\def\D{\Delta}

\def\L{\Lambda}
\def\w{\omega}
\def\th{\theta}
\def\bz{{\bar z}}

\def\bw{{\bar w}}

\def\CA{{\mathcal A}}

\def\CC{{\mathcal C}}
\def\CD{{\mathcal D}}

\def\CH{{\mathcal H}}
\def\CI{{\mathcal I}}

\def\CN{{\mathcal N}}
\def\CO{{\mathcal O}}

\def\CR{{\mathcal R}}

\def\Tr{{\text{Tr}}}

\newcommand{\dt}{{\text d}}
\newcommand{\p}{\partial}

\newcommand{\la}{{\langle}}
\newcommand{\ra}{{\rangle}}
\newcommand{\eff}{{\text{eff}}}

\def\wh{\hat}
\def\uv{{\text{\begin{tiny}UV\end{tiny}}}}
\def\sho{{\text{\begin{tiny}SHO\end{tiny}}}}

\NewDocumentCommand{\codeword}{v}{%
\texttt{\textcolor{blue}{#1}}%
}

\usepackage{subfiles} 

\begin{document}


\title{Effective density matrix for vacua in asymptotically flat gravity}
\author{Temple He}
\affiliation{Walter Burke Institute for Theoretical Physics, California Institute of Technology, Pasadena, CA 91125}
\author{Prahar Mitra}
\affiliation{Institute for Theoretical Physics, University of Amsterdam, Science Park 904, Postbus 94485, 1090 GL Amsterdam, The Netherlands}
\author{Kathryn M. Zurek}
\affiliation{Walter Burke Institute for Theoretical Physics, California Institute of Technology, Pasadena, CA 91125}

\begin{abstract}
\noindent We explicitly construct the density matrix associated to the vacuum state of a large spherically symmetric causal diamond of area $A$ in four-dimensional asymptotically flat gravity. We achieve this using the soft effective action, which characterizes the low-energy gravitational degrees of freedom that arise in the long-distance limit of the Einstein--Hilbert action and consists of both the soft graviton mode and the Goldstone mode arising from the spontaneous breaking of supertranslation symmetry. Integrating out the soft graviton mode, we obtain an effective action for purely the Goldstone mode, from which we extract the density matrix and therefore the modular Hamiltonian $\hat K_s$ associated to the vacuum state. As a corollary, we explicitly compute the mean and variance of $\hat K_s$, finding $\langle \Delta {\hat K}_s^2 \rangle = A/\epsilon_{\text{\begin{tiny}UV\end{tiny}}}^2$, with $\epsilon_{\text{\begin{tiny}UV\end{tiny}}}$ being a length-scale UV cutoff on the celestial sphere.
\end{abstract}

\maketitle

\preprint{CALT-TH 2025-030}

\maketitle
	
\noindent {\bf Introduction.}  The description of subregions of spacetime in quantum gravity has been an ongoing focus of intense interest \cite{Hamilton:2005ju, Ryu:2006bv, Jacobson:2015hqa, Jacobson:2018ahi, Verlinde:2019ade, Giddings:2021khn, Leutheusser:2022bgi, Jensen:2023yxy, Ciambelli:2024swv, Bub:2024nan, Caminiti:2025hjq, Ciambelli:2025flo, Ciambelli:2025ztm, Fransen:2025npa}. Such subregions generically arise in the presence of horizons, which by their nature divide spacetime into separate regions. Due to the presence of soft, or low-energy, long-wavelength modes in gravity that cross subregions, the description of spacetime typically requires degrees of freedom that propagate on the horizon. Recently, it has been shown that such ``edge'' modes play the central role in the description of gravity (and gauge theories) in a subregion~\cite{Donnelly:2014fua, Donnelly:2015hxa, David:2022jfd, Chen:2023tvj, He:2024ddb, Chen:2024kuq, Ball:2024hqe, He:2024skc, Ball:2024xhf, Ball:2024gti, Araujo-Regado:2024dpr}. 

The soft effective action (SEA), introduced in \cite{Kapec:2021eug}, is a $d$-dimensional action describing the edge modes~\cite{Chen:2023tvj, He:2024ddb, Chen:2024kuq, Ball:2024hqe, He:2024skc, Araujo-Regado:2024dpr} localized on a codimension-2 surface, e.g., the bifurcate horizon of a causal diamond or the past or future boundary of spatial infinity of Minkowski spacetime, in a $(d+2)$-dimensional gauge theory or gravity. The original construction of the SEA in \cite{Kapec:2021eug} for both gauge theory and gravity uses basic ideas from effective field theory and symmetry breaking to describe the infrared factorization of scattering amplitudes, namely soft theorems and infrared divergences (see also \cite{Himwich:2020rro, Nguyen:2021ydb, Kalyanapuram:2021tnl} for earlier related work). In particular, for gravity, the SEA is an action consisting of the leading soft graviton mode $\CN$ as well as the Goldstone mode $\CC$ arising from the spontaneously broken supertranslation symmetry. If the Goldstone mode is non-dynamical, the tree-level SEA was recently shown to be equivalent to the supertranslation charge \cite{He:2014cra}, as well as the effective action for shockwaves~\cite{He:2024vlp}. Furthermore, though not constructed originally with subregions or their edge modes in mind, in the context of abelian gauge theories, the corresponding tree-level SEA, in addition to being equivalent to the large gauge charge, was further shown to be equivalent to soft Wilson line dressings~\cite{He:2024ddb, He:2024skc}, implying a deep connection to non-local observables arising in the presence of horizons \cite{Araujo-Regado:2024dpr, Giddings:2025xym}.

 \begin{figure}[t]
\centering
\tdplotsetmaincoords{85}{115} 
\begin{tikzpicture}[tdplot_main_coords, scale=1.5]

\draw[->] (5,0,0.5) -- (5,0,1) node[above] {$t$};

\def\a{2.5} 
\def\b{1.5} 

\draw[thick,red,fill=red!20,opacity=0.7] (-\a,-\b,0) -- (\a,-\b,0) -- (\a,\b,0) -- (-\a,\b,0) -- cycle;

\def\R{1} 
\def\H{1} 

\shade[gray , thick, shading=radial, inner color=gray!40, outer color=gray!80] (0,0,0) circle (\R);
\draw[thick,blue] (0,0,0) circle (\R);

\node at (0,1.3,0.1) { {\color{blue} $S^2$} };

\foreach \angle in {0,30,60,90,120,-30,-60}
{
\draw[] (0,0,\H) -- ({\R*cos(\angle)}, {\R*sin(\angle)}, 0);
\draw[] (0,0,-\H) -- ({\R*cos(\angle)}, {\R*sin(\angle)}, 0);
}   

\foreach \angle in {155,180,210,240,260}
{
\draw[dashed] (0,0,\H) -- ({\R*cos(\angle)}, {\R*sin(\angle)}, 0);
\draw[dashed] (0,0,-\H) -- ({\R*cos(\angle)}, {\R*sin(\angle)}, 0);
}

\end{tikzpicture}
\caption{\small A causal diamond in asymptotically flat spacetime. We associate to it a soft density matrix ${\wh \rho}_s$, which is obtained by tracing out the energetic modes, leaving only the edge mode degrees of freedom on the blue $S^2$. We will also be interested in ${\wh K}_s \equiv -\ln {\wh \rho}_s$, which is the associated modular Hamiltonian in the soft Hilbert space.} 
\label{fig:CD}
\end{figure}  

To construct the density matrix ${\wh \rho}_s$ associated to the vacuum state of a large causal diamond of radius $R$, whose boundary consists of ingoing and outgoing null rays, in asymptotically flat spacetime, we will utilize the SEA. Indeed, in the limit where the causal diamond is large compared to the Planck length, i.e., $R \gg \ell_P$, the edge modes parametrizing the area fluctuations of the causal diamond are well-approximated by the soft (low-energy) modes in asymptotically flat spacetimes, and their dynamics is described by the SEA \cite{He:2024ddb, Araujo-Regado:2024dpr}.\footnote{Strictly speaking, \cite{He:2024ddb, Araujo-Regado:2024dpr} focus on abelian gauge theory rather than gravity. To truly demonstrate this approximation holds in gravity, we would need to show how the SEA arises from the Einstein--Hilbert action by taking a large causal diamond limit, and we reserve this for future work.} 

By viewing the causal diamond as an open quantum system, the vacuum density matrix ${\wh \rho}_s$, which describes the soft sector, is obtained by tracing out all the energetic modes, leaving only the edge mode degrees of freedom on the bifurcate horizon (see Figure~\ref{fig:CD}). In particular, we will show in this letter that there exists a ${\wh \rho}_s$ such that 
\begin{equation}
\begin{split}
\label{rho-criterion}
\la {\wh \CO}_1 \cdots {\wh \CO}_n \ra &\equiv \Tr_{\CH_s}\big( {\wh \rho}_s {\wh \CO}_1 \cdots {\wh \CO}_n \big)  \\
&= \int[ \CD \CC] e^{-S_\eff[\CC]} \CO_1 \cdots \CO_n , 
\end{split}
\end{equation}
where ${\wh \CO}_i$ are operator insertions involving the Goldstone, whereas $\CO_i$ are the corresponding field insertions in the path integral, and the trace is over the soft sector of the Hilbert space $\CH_s$ of general relativity. 

Having computed the soft density matrix ${\wh \rho}_s$, one can proceed to compute many possible information-theoretic quantities of interest. In this letter, we focus on computing both the mean and variance of the soft modular Hamiltonian ${\wh K}_s \equiv - \ln {\wh \rho}_s$. The entanglement entropy, given by the mean $\la {\wh K}_s \ra$, has garnered much attention in the past, and in gravity it has been established to obey an area law \cite{PhysRevD.34.373, Srednicki:1993im, Ryu:2006ef, Ryu:2006bv, Eisert:2008ur, Casini:2011kv, Brauer:2013lta, Balasubramanian:2023dpj}. On the other hand, fluctuations of the modular Hamiltonian, namely $\D {\wh K}_s \equiv {\wh K}_s - \la {\wh K}_s \ra$, are much less studied. Recently, it has been argued that such modular fluctuations of a causal diamond of area $A$ appear to satisfy the relation 
\begin{equation}
\begin{split}
\label{mod-fluc}
\la \D {\wh K}_s^2 \ra = \la {\wh K}_s \ra=\frac{A}{\e_\uv^2} ,
\end{split}
\end{equation}
where $\e_\uv$ is a length-scale UV cutoff. This was verified using the AdS/CFT correspondence \cite{Perlmutter:2013gua, Verlinde:2019ade, DeBoer:2018kvc}, in Minkowski spacetime \cite{Verlinde:2019xfb, Banks:2021jwj, Verlinde:2022hhs, He:2024vlp, Fransen:2025npa}, and in cosmological settings \cite{Aalsma:2025bcg}. It was more precisely shown in \cite{Verlinde:2019ade, Fransen:2025npa} that, at least in the setups considered therein, $\e_\uv^2 = 4G_N$. Importantly, Eq.~\eqref{mod-fluc} implies that the size of modular fluctuations is sensitive to the IR scale controlled by $A$, as well as the UV scale controlled by $\e_\uv$. Utilizing the density matrix constructed from the soft modular Hamiltonian, we are able to verify that while $\la \D \hat K_s^2 \ra$ is precisely that given in Eq.~\eqref{mod-fluc}, $\la \hat K_s \ra$ receives a logarithmic correction arising from the non-normalizability of the $\hat K_s$-eigenstates. However, since previous results obtaining Eq.~\eqref{mod-fluc} ignored loop corrections and relied on a purely semiclassical analysis, which we do not expect to be sensitive to the logarithmic correction, our results are consistent with those in the literature.  We leave for future work the precise implementation of an appropriate normalization for the eigenstates, although we will comment below on some physically motivated possibilities. 

Indeed, the structure of Eq.~\eqref{mod-fluc} follows straightforwardly from the structure of the soft density matrix ${\wh \rho}_s$, which we will show to have a Gaussian distribution of the soft Goldstone mode given by (see Eq.~\eqref{soft-density-matrix})
\begin{equation}\label{eq:intro-rho}
{\wh \rho}_s = e^{ - {\wh K}_s } \sim \exp\bigg( - c \oint_z {\wh \CC}^{zz} {\wh \CC}_{zz} \bigg),
\end{equation} 
where $c$ is a dimensionful constant we compute below. This density matrix implies that there is a non-trivial distribution of $\hat \CC$-vacua for a large causal diamond in asymptotically flat spacetimes. This is in stark contrast to the trivial vacuum structure usually assumed in a quantum field theory, and clarifies the fact that $\la \D \hat K_s^2 \ra$ involves the IR scale because of nontrivial entanglement present in the vacuum. Indeed, such an effect is difficult to capture with traditional field theory techniques, where the vacuum in perturbative gravity is assumed to be unique. Furthermore, if the entanglement entropy counts the number of gravitational qudits $N$ in the vacuum state of the causal diamond, such that $N \equiv \langle \hat K_s \rangle$, then the Gaussian distribution of $\hat \CC$-vacua implies that fluctuations obey ``root-$N$'' statistics, i.e., $\la \Delta \hat K_s ^2 \ra^{1/2} \sim \sqrt N$. We leave the study of vacuum transitions associated to this structure, and their implications for spacetime, to future work.

\medskip

\noindent{\bf Asymptotically Flat Metrics.} 
To describe four-dimensional asymptotically flat spacetimes, we work in retarded ($+$) or advanced ($-$) Bondi-Sachs coordinates $(u_\pm,r,z_\pm,\bz_\pm)$. In these coordinates, past and future null infinity $\CI^\pm$ are located at $r \to \infty$ while keeping the other coordinates fixed. Moreover, $u_\pm$ is the future-directed null coordinate along the generators of $\CI^\pm$, and $(z_\pm,\bz_\pm)$ are stereographic coordinates on the transverse cut of $\CI^\pm$, which we assume has the $S^2$-topology. We choose coordinates so that the past and future transverse coordinates are antipodally matched, i.e., $z_+|_{\CI^+} = - 1/\bz_-|_{\CI^-}$.\footnote{On $\CI^\pm$, the stereographic coordinates $z_\pm$ are related to the standard angular coordinates $(\th,\phi)$ via $z_+ =e^{i\phi} \tan \frac{\t}{2}$, and $z_- =e^{i(\pi+\phi)} \tan \frac{\pi-\t}{2}$.} The past (future) boundary of $\CI^+$ ($\CI^-$) is located at $u_+=-\infty$ ($u_-=+\infty$) and is denoted by $\CI^+_-$ ($\CI^-_+$). While we will denote fields on $\CI^\pm$ with a superscript $\pm$, we will for notational clarity drop the $\pm$ subscript on the coordinates. 

Near $\CI^\pm$, asymptotically flat metrics take the form
\begin{equation}
\begin{split}
\label{metric-exp}
	\dt s^2 &= - \dt u^2 \mp 2 \dt u \dt r + 2 r^2 \g_{z\bz} \dt z \dt \bz \\
&\qquad + \frac{2m^\pm_B}{r} \dt u^2 + r C^\pm_{zz} \dt z^2 + r C^\pm_{\bz\bz} \dt \bz^2 + \cdots ,
\end{split}
\end{equation}
where $\g_{z\bz} = 2(1+ z\bz)^{-2}$ is the unit round metric on the celestial sphere $S^2$, and $\cdots$ denotes subleading terms in the large $r$ expansion. The news tensor $N^\pm_{zz} \equiv \p_u C^\pm_{zz}$ characterizes outgoing (on $\CI^+$) or incoming (on $\CI^-$) gravitational radiation, and $m^\pm_B$ describes the local energy density of the radiation. To describe the low-energy sector of asymptotically flat spacetimes, we define the fields\footnote{These modes are well-defined in spacetimes studied by Christodoulou and Klainerman in \cite{Christodoulou:1993uv}. The results here can be extended to more general spacetimes (including ones where the peeling theorem fails) with minimal modifications.}
\begin{equation}
\begin{split}
\label{CK-constraint}
\CC^\pm_{zz} &\equiv C_{zz}^\pm \big|_{u=\mp\infty} = - 2 D^2_z \CC^\pm , \\
\CN^\pm_{zz} &\equiv \int_{\mathbb R} \dt u N^\pm_{zz} = - 2 D_z^2  \CN^\pm ,
\end{split}
\end{equation}
where $D_z$ is the $\g$-covariant derivative, and $\CN^\pm$ are the leading soft graviton modes, and $\CC^\pm$ are the supertranslation Goldstone modes on $\CI^\pm$. Notice that because the physical degrees of freedom are $\CC_{zz}^\pm$ and $\CN_{zz}^\pm$, any mode in $\CC^\pm$ and $\CN^\pm$ that is annihilated by $D_z^2$ and $D_\bz^2$ is pure gauge. This means we have the gauge symmetry
\begin{equation}
\begin{split}
\label{gauge-freedom}
\CC^\pm \sim \CC^\pm + f_{\text{gauge}} , \qquad \CN^\pm \sim \CN^\pm + f_{\text{gauge}} ,
\end{split}
\end{equation}
where $D_z^2 f_{\text{gauge}} = D_\bz^2 f_{\text{gauge}} = 0$. Its explicit form is given by
\begin{align}\label{gauge}
f_{\text{gauge}} = \frac{a + b z + {\bar b} \bz+cz\bz}{1+z\bz} ,
\end{align}
where $a,c\in {\mathbb R}$ and $b \in {\mathbb C}$, thus implying the gauge group is ${\mathbb R}^4$.

Fields on $\CI^+_-$ and $\CI^-_+$ satisfy an antipodal matching condition\footnote{The Bondi mass aspect is related to the news tensor in the vacuum via the constraint equation
\begin{align*}
\p_u m_B^\pm = \frac{1}{4} \big( D_z^2 N^{\pm zz} + D_\bz^2 N^{\pm\bz\bz} \mp N_{zz}^\pm N^{\pm zz} \big).
\end{align*}
}
\begin{equation}
\begin{split}
\label{matching-cond}
\CC^+ = -  \CC^- , \qquad m_B^+ \big|_{\CI^+_-} = m_B^- \big|_{\CI^-_+}  .
\end{split}
\end{equation}
Together, we will refer to $\CC^\pm$ and $\CN^\pm$ as the \emph{soft modes}, and these soft modes are the edge modes of our causal diamond of radius $R$. Upon quantization, the soft modes $\CC^\pm$ and $\CN^\pm$ are elevated to operators $\wh\CC^\pm$ and $\wh\CN^\pm$ on soft Hilbert spaces $\CH^\pm_s$.\footnote{Including the radiative (hard) degrees of freedom, the full $out$ and $in$ Hilbert spaces are $\CH^\pm=\CH_s^\pm \otimes \CH_r^\pm$. While $\CH^+$ is isomorphic to $\CH^-$ with the isomorphism is given by the $S$-matrix, the soft Hilbert spaces by themselves are not isomorphic.} Their commutators can be determined via the covariant phase space formalism \cite{Crnkovic:1986ex, Wald:1999wa, Harlow:2019yfa, He:2020ifr, He:2023bvv}, which are given by
\begin{align}
\label{soft-comm}
\left[ {\wh \CC}_{zz}^\pm(z,\bz) ,  {\wh \CN}^{\pm}_{\bw\bw} (w,\bw) \right] &=  \frac{i\ka^2}{2} \g_{z\bz} \d^2(z-w) ,
\end{align}
where $\ka \equiv \sqrt{32\pi G_N}$. Because ${\wh \CC}^\pm$ commutes with itself, eigenstates of ${\wh \CC}^\pm$ form an orthogonal basis on $\CH_s^\pm$, so that
\begin{equation}
\begin{split}
\label{C-eigenstates}
{\wh \CC}^\pm(z,\bz) |\CC,\pm\ra &= \CC(z,\bz) |\CC,\pm\ra , \\
\la \CC, \pm |\CC' , \pm \ra &= \delta(\CC-\CC') . 
\end{split}
\end{equation}
Traces in the soft Hilbert space can then be evaluated using this eigenbasis to be
\begin{equation}
\begin{split}
\label{trace-def}
\Tr_{\CH^\pm_s} [ \CO ]  \equiv \int [ \CD \CC ] \la \CC,\pm | \CO  | \CC,\pm \ra,
\end{split}
\end{equation}
where the measure $[\CD \CC]$ is defined so that
\begin{equation}
\begin{split}
\label{measure-normalization}
\int [\CD \CC ] \d ( \CC - \CC' ) = 1 . 
\end{split}
\end{equation}

Let us now turn to the construction of the SEA. It was shown in \cite{Bondi:1962px, Sachs:1962wk, Strominger:2013jfa} that asymptotically flat spacetimes admit an infinite-dimensional symmetry that acts on the soft modes via 
\begin{equation}
\begin{split}
\label{supertranslations}
{\wh \CC}^\pm \to {\wh \CC}^\pm \pm f , \qquad {\wh \CN}^\pm \to {\wh \CN}^\pm  .
\end{split}
\end{equation}
where $f(z,\bz)$ is any smooth function on the celestial sphere. This symmetry, known as BMS supertranslations, is related to Weinberg's soft theorem \cite{Weinberg:1965aa, He:2014laa}, infrared divergences \cite{Yennie:1961ad, Kapec:2017tkm}, and memory effects \cite{PhysRevD.45.520, Strominger:2014pwa}. The symmetry is spontaneously broken by the gravitational vacuum, and the dynamics of the corresponding Goldstone mode\footnote{If $f=f_\text{gauge}$, then Eq.~\eqref{supertranslations} is a trivial transformation due to Eq.~\eqref{gauge-freedom}. Consequently, ${\wh \CC}$ is the Goldstone mode for the symmetry breaking of $\text{supertranslations}/{\mathbb R}^4$.}  ${\wh \CC} \equiv {\wh \CC}^+ = -  {\wh \CC}^-$ and the soft graviton mode ${\wh \CN} \equiv {\wh \CN}^+ + {\wh \CN}^-$ was shown in \cite{Kapec:2021eug} to be described by the SEA\footnote{In \cite{Kapec:2021eug}, the authors worked with the flat metric on $S^2$. We have rewritten their result for the case of the round metric instead. Furthermore, we use a different slightly definition for $\a$ than that found in \cite{Kapec:2021eug} in that we scale out $G_N$.}
\begin{equation}
\begin{split}
\label{sea0}
S_{\text{KM}} [\CC,\CN] &= \frac{2}{\ka^2} \oint_z \left( \frac{\a}{2\pi} \CN^{zz} \CN_{zz}  - i \CC^{zz} \CN_{zz} \right),
\end{split}
\end{equation}
where
\begin{equation}
\begin{split}
\oint_z \equiv \int_{S^2} \dt^2 z \g_{z\bz} , \qquad \a \equiv \ln \frac{\L}{\mu} . 
\end{split}
\end{equation}
Here, $\mu$ is an infrared cutoff that defines the four-dimensional theory, and $\L$ is a cutoff that delineates soft modes from hard modes, in that soft modes have energy below $\L$ and hard modes have energy above $\L$.\footnote{While $\a$ is also related to the radius $R$ of the causal diamond associated to the action in Eq.~\eqref{sea0}, the precise relationship will not be relevant for our work.} Importantly, \cite{Kapec:2021eug} proves that the SEA reproduces the low-energy factorization properties of four-dimensional scattering amplitudes.  

For our purposes, we will only consider correlators of $\CC$, meaning we can integrate out $\CN$. We evaluate the equation of motion for $\CN$ by varying Eq.~\eqref{sea0} and find
\begin{align}\label{N-eom}
\CN_{zz} = \frac{i\pi}{\a} \CC_{zz}.
\end{align}
Substituting this into Eq.~\eqref{sea0}, we obtain the effective action for $\CC$:
\begin{equation}
\begin{split}
\label{sea}
S_{\text{eff}} [\CC] &= \frac{\pi}{\a\ka^2} \oint_z \CC^{zz} \CC_{zz} . 
\end{split}
\end{equation}
We now proceed to construct a density matrix $\rho_s$ such that Eq.~\eqref{rho-criterion} is satisfied.

\medskip

\noindent{\bf A Soft Density Matrix.} We propose that the density matrix associated to the vacuum state of the causal diamond is given by\footnote{The formally divergent quantity $\ln \la \CC | \CC \ra$, which arises because the ${\wh \CC}$-eigenstates are not normalizable, is addressed after Eq.~\eqref{final}.}
\begin{equation}
\begin{split}
\label{soft-density-matrix}
{\wh \rho}_s \equiv e^{ - S_{\text{eff}} [{\wh \CC}] - \ln \la \CC | \CC \ra } ,
\end{split}
\end{equation}
where we have dropped the $\pm$ label on the ${\wh \CC}$-eigenstates owing to the first matching condition in Eq.~\eqref{matching-cond}. Using Eqs.~\eqref{trace-def} and \eqref{soft-density-matrix}, and assuming that ${\wh \CO}_k \equiv \CO_k({\wh \CC})$, it follows that
\begin{equation}
\begin{split}
\label{soft-trace}
\Tr_{\CH_s}\big({\wh \rho}_s {\wh \CO}_1 \cdots {\wh \CO}_n\big) &= \int [ \CD \CC ] e^{ - S_{\text{eff}} [\CC] } \CO_1 \cdots \CO_n,
\end{split}
\end{equation}
which is precisely Eq.~\eqref{rho-criterion}.\footnote{For operators that depend on ${\wh \CN}$, we can use Eq.~\eqref{N-eom} to replace ${\wh \CN} \to \frac{i\pi}{\a} {\wh \CC}$. Up to contact terms, this correctly reproduces insertions of ${\wh \CN}$.} To ensure the density matrix ${\wh \rho}_s$ is properly normalized so that its trace is unity, we require
\begin{equation}
\begin{split}
\label{measure-constraint}
\Tr_{\CH_s}\big({\wh \rho}_s\big) = \int [ \CD \CC ] e^{ - S_{\text{eff}} [\CC] }  = 1.
\end{split}
\end{equation}
In the Supplementary Materials, we provide an explicit construction of a measure $[\CD \CC]$ that satisfies Eq.~\eqref{measure-constraint}.

Next, we compute the mean and variance of the modular Hamiltonian ${\wh K}_s \equiv - \ln {\wh \rho}_s$, which using Eq.~\eqref{soft-density-matrix} is given by
\begin{align}\label{K-def}
\begin{split}
{\wh K}_s &\equiv S_{\text{eff}} [{\wh \CC}] + \ln \la \CC | \CC \ra . 
\end{split}
\end{align}
As we shall see, the fact the SEA in Eq.~\eqref{sea} is quadratic implies that our density matrix obeys Gaussian statistics. To compute its mean, we note that
\begin{equation}
\begin{split}
\label{K_mean_1}
\la {\wh K}_s \ra &= \Tr_{\CH_s} \big( {\wh \rho}_s {\wh K}_s \big) \\
&= \int [ \CD \CC ] e^{ - S_{\text{eff}} [\CC] } S_{\text{eff}} [\CC] + \ln \la \CC | \CC \ra .
\end{split}
\end{equation}
The path integral appearing can be evaluated by expanding $\CC$ in a basis of real spherical harmonics $Y_{\ell m}$ on $S^2$, given by
\begin{equation}
\begin{split}
\label{C-decomp}
\CC(z,\bz) &= \ka \sum_{\ell=0}^{\ell_{\max}} \sum_{m=1}^{2\ell+1} \CC_{\ell m} Y_{\ell m}(z,\bz) , \\
Y_{\ell m}^* &= Y_{\ell m} , \qquad \oint_z Y_{\ell m} Y_{\ell' m'} = \d_{\ell\ell'} \d_{mm'} ,
\end{split}
\end{equation}
where the factor of $\ka$ is introduced so that the coefficients $\CC_{\ell m}$ are dimensionless, and we have restricted $\ell \leq \ell_{\max}$ to regulate UV divergences in the path integral. To relate $\ell_{\max}$ to a UV length-scale $\e$, we observe that the total number of constants required to specify the function $\CC$ in Eq.~\eqref{C-decomp} is given by $(\ell_{\max}+1)^2$. Alternatively, we can regulate the UV divergences by breaking up the celestial sphere into plaquettes of area $\e^2$. The number of constants required to describe $\CC$ is then $A/\e^2$, where $A=4\pi R^2$ is the area of the celestial sphere.\footnote{By fixing a finite radius $R$, we have explicitly broken large diffeomorphism invariance. This is likely equivalent to choosing a supertranslation frame associated to a particular observer \cite{Kapec:2016aqd}.} Clearly, these two numbers should be equal, which implies 
\begin{equation}
\begin{split}
\label{Lmax_identification}
(\ell_{\max}+1)^2 = \frac{A}{\e^2} \quad \implies \quad \ell_{\max} = \sqrt{\frac{A}{\e^2}} - 1 .
\end{split}
\end{equation}
Using Eq.~\eqref{C-decomp}, we can evaluate Eq.~\eqref{sea} to be
\begin{align}\label{ortho-prod}
\begin{split}
S_\eff[\CC] &= \frac{4\pi}{\a} \sum_{\ell,m,\ell',m'} \!\!  \CC_{\ell m} \CC_{\ell'm'} \oint_z D_z^2 Y_{\ell m} (D^z)^2 Y_{\ell'm'} \\
&= \frac{\pi}{\a} \sum_{\ell=2}^{\ell_{\max}}\sum_{m=1}^{2\ell+1} (\ell-1) \ell ( \ell + 1 ) ( \ell + 2 ) \CC_{\ell m}^2 ,
\end{split}
\end{align}
where we used the fact that $Y_{\ell m}$ are eigenstates of the transverse Laplacian $\Box \equiv D_z D^z + D^z D_z$ with eigenvalue $-\ell(\ell+1)$. Note that the $\ell = 0,1$ spherical harmonics of $\CC$ do not appear in the action, which is expected since they are precisely the pure gauge modes in $\CC$ (see Eq.~\eqref{gauge-freedom}).

We can now perform the integral in Eq.~\eqref{K_mean_1} using the precise definition of the measure given by Eq.~\eqref{ansatz} in the Supplementary Materials. The path integral is now simply a series of Gaussian integrals, and we find
\begin{equation}
\begin{split}
\label{eq:Ks}
\la {\wh K}_s \ra &= \frac{1}{2} (\ell_{\max}+3)(\ell_{\max}-1) + \ln \la \CC | \CC \ra \\
&= \frac{1}{2} \left( \frac{A}{\e^2} - 4 \right) + \ln \la \CC | \CC \ra . 
\end{split}
\end{equation}
In the regime of interest, namely $A/\e^2 \gg 1$, we can ignore the $-4$ term, which is present due to the gauge symmetry in Eq.~\eqref{gauge-freedom}.

We next turn to the variance of ${\wh K}_s$, which is defined to be
\begin{equation}
\begin{split}
\la \D {\wh K}_s^2 \ra &\equiv \la {\wh K}_s^2 \ra - \la {\wh K}_s \ra^2  \\
&= \int [ \CD \CC ] e^{ - S_{\text{eff}} [\CC] }  S_{\text{eff}} [\CC]^2  -  \la S_{\text{eff}} [{\wh \CC}] \ra^2 .
\end{split}
\end{equation}
The path integral in the first term is similarly straightforward to evaluate, and the second term was evaluated previously in Eq.~\eqref{eq:Ks}. It follows the final result is
\begin{align}
\label{final}
\begin{split}
\la \D {\wh K}_s^2 \ra &= \frac{1}{2} \left( \frac{A}{\e^2} - 4 \right) , 
\end{split}
\end{align}
which is precisely of the form given in Eq.~\eqref{mod-fluc}. Comparing with Eq.~\eqref{eq:Ks}, we see that except for the term $\ln \la \CC | \CC \ra$, $\la \D {\wh K}_s^2 \ra$ is precisely equal to $\la {\wh K}_s \ra$. Similar results for the modular Hamiltonian of a causal diamond have previously been obtained in \cite{Verlinde:2019ade, Banks:2021jwj, Verlinde:2022hhs}. 

Finally, let us now discuss the last term in Eq.~\eqref{eq:Ks}, which may appear troubling since it is divergent. This divergence arises from the fact that the eigenstates of $\hat\CC$ are non-normalizable, which arises due to the fact that the conjugate of ${\wh \CC}$, namely ${\wh \CN}$, is non-compact. The same divergence arises in the thermal density matrix of a free particle on a line, and indeed, our density matrix in Eq.~\eqref{soft-density-matrix} is qualitatively similar to the free particle Hamiltonian. To be precise, we show in the Supplementary Materials that for ${\wh H} = \frac{{\wh P}^2}{2m}$ with ${\hat X} \in {\mathbb R}$, the mean of the modular Hamiltonian is (see Eq.~\eqref{KT-line_app})
\begin{equation}
\begin{split}
\label{KT-line}
\la {\hat K}_\text{line}(\b) \ra = \frac{1}{2}  + \ln \left( \sqrt{ \frac{m}{2\pi\b}} \la p | p \ra \right) . 
\end{split}
\end{equation}

The divergence in Eq.~\eqref{KT-line} due to the logarithm of $\la p|p\ra$ is precisely the same one in Eq.~\eqref{eq:Ks} and can be regulated in several ways.  We explore two such ways in the Supplementary Materials. For instance, if we compactify ${\hat X} \sim {\hat X} + 2\pi R$, then we show that for large $R$, we have
\begin{equation}
\begin{split}
\label{KT-ring}
\la {\hat K}_\text{ring}(\b) \ra ~ \xrightarrow{R \to \infty} ~ \frac{1}{2}  + \ln \left( \sqrt{ \frac{m}{2\pi\b}} (2\pi R) \right), 
\end{split}
\end{equation}
which regulates $\la p | p \ra$  in Eq.~\eqref{KT-line} to $2\pi R$. Alternatively, the divergence can be regulated by turning on a small potential for the field.  For example, we consider a simple harmonic oscillator described by the Hamiltonian ${\wh H} = \frac{{\wh P}^2}{2m} + \frac{1}{2} m \o^2 {\wh X}^2$. In this case, we find that for small $\o$,
\begin{equation}
\begin{split}
\label{KT-sho}
\la {\wh K}_\sho(\b) \ra ~ \xrightarrow{\o \to 0} ~ 1 - \ln ( \b \o ) ,
\end{split}
\end{equation}
which regulates $\la p | p \ra $ in Eq.~\eqref{KT-line} to $\sqrt{ 2\pi/(m\o) }$.

Indeed, the latter type of regulation is precisely what we expect to occur for the SEA as well once we allow the soft graviton and Goldstone modes to have small but nonzero energy. More precisely, the SEA in Eq.~\eqref{sea0} describes the dynamics of soft modes ${\wh \CC}$ and ${\wh \CN}$, which have exactly zero energy ($\o=0$), while the holographic model that describes all of four-dimensional gravity should include modes of all energies $\o > 0$. At lowest derivative order, we expect the effective action of the mode at energy $\o$ will be precisely that of a harmonic oscillator. It follows the mean and variance of the modular Hamiltonian can be obtained (at least to lowest derivative order) by integrating $\la {\wh K}_\sho(\b) \ra$ over $\o$ with a frequency density $n(\o)$ that depends on the details of four-dimensional theory. As we show in the Supplementary Materials, if we assume a uniform distribution (which is expected at lowest derivative order), we obtain the equality 
\begin{align}
\la {\wh K}_\sho(\b) \ra = \la \D {\wh K}^2_\sho(\b) \ra,
\end{align}
a relation that has been obtained in a variety of setups \cite{Verlinde:2019xfb, Banks:2021jwj, Verlinde:2022hhs, He:2024vlp, Fransen:2025npa}. Thus, to summarize, we view the logarithmic divergence in Eq.~\eqref{eq:Ks} as a symptom of the zero energy limit, which is the realm where the SEA is applicable. We leave an analysis on how to appropriately uplift the SEA to nonzero energies for future work.

Finally, we observe that the scaling of $\la \D {\wh K}_s^2 \ra$ as the area in Eq.~\eqref{final} was also obtained using other arguments for flat spacetimes \cite{Verlinde:2019xfb, Banks:2021jwj, Verlinde:2022hhs, He:2024vlp, Fransen:2025npa}. However, whereas all such previous arguments involved either thermodynamic arguments or were effectively a semiclassical analysis of perturbative gravity, Eq.~\eqref{final} takes into account quantum effects coming from the IR divergent term in the SEA, demonstrating that the scaling survives loop corrections.

\medskip

\noindent{\bf Conclusions.} In this letter, we have shown that by taking the soft effective action derived in \cite{Kapec:2021eug} as our starting point, we are able to construct a density matrix ${\wh \rho}_s$ in the soft Hilbert space of a large causal diamond that reproduces both the leading soft theorem as well as the IR factorization found in $S$-matrix elements. This density matrix captures the entanglement in the vacuum sector of gravitational theories in asymptotically flat spacetimes. By further computing the variance of the modular Hamiltonian, we are able to quantify the fluctuation of the entanglement and show that it obeys an area law. In fact, because we have constructed the density matrix, we can further compute other information-theoretic quantities that are of interest.

We would like to make a few comments regarding the density matrix that we have constructed. In four spacetime dimensions, the parameter $\a$ is logarithmically divergent (as $\mu/\L \to 0$) and captures precisely the IR divergences in the one-loop determinant of the graviton. Hence, if we formally take this limit inside the integrand in Eq.~\eqref{soft-density-matrix}, we see that the density matrix becomes proportional to the identity. This suggests that the density matrix in the vacuum sector has a flat entanglement spectrum and is maximally mixed. This result was obtained in the context of de Sitter spacetime in \cite{Dong:2018cuv, Chandrasekaran:2022cip}, and more generally in the context of maximally symmetric spacetimes in \cite{Jacobson:2015hqa}.

On the other hand, if we are interested in phenomena that depends sensitively on the IR data, we do not want to take $\a \to \infty$. Instead, $\a$ is formally a large, but finite, number. In this case, the density matrix is no longer maximally mixed. As was argued in \cite{Banks:2022irh}, again in the context of de Sitter spacetime, one does not need a flat entanglement spectrum to obtain Eq.~\eqref{final}. Rather, it holds for any value of $\a$.

Finally, we conclude by observing that Eq.~\eqref{K-def} implies
\begin{align}
\begin{split}
\D {\wh K}_s^2 &\equiv ({\wh K}_s - \la {\wh K}_s \ra)^2 \\
&= \bigg( \frac{\pi}{\a\ka^2} \oint_{z} {\wh \CC}^{zz} {\wh \CC}_{zz}  - \la S_{\eff}[{\wh \CC}]\ra \bigg)^2,
\end{split}
\end{align}
which means that $\D {\wh K}_s^2$ scales as the four-point function of the Goldstone mode $\CC$. As $\CC$ is part of the physical data present in the shear $C_{zz}$, this suggests that modular fluctuations, which have the same scaling behavior as area fluctuations, involve the four-point function of length fluctuations in a perturbative quantum theory of gravity. It would be of great interest to try to connect such length fluctuations to actual observables within an experimental setup, and we hope to address this in future work.

\medskip

\noindent\textbf{Acknowledgments.}
We would like to thank Daniel Kapec, Cynthia Keeler, and Ronak Soni for useful discussions. This work was performed in part at Aspen Center for Physics, which is supported by National Science Foundation grant PHY-2210452. T.H. and K.Z. are supported by the Heising-Simons Foundation “Observational Signatures of Quantum Gravity” collaboration grant 2021-2817, the U.S. Department of Energy, Office of Science, Office of High Energy Physics, under Award No. DE-SC0011632, and the Walter Burke Institute for Theoretical Physics. P.M. is supported by the European Research Council (ERC) under the European Union’s Horizon 2020 research and innovation programme (grant agreement No 852386). K.Z. is also supported by a Simons Investigator award.

\bibliography{references_use.bib}

\clearpage
\onecolumngrid
\begin{center}
\textbf{\large SUPPLEMENTARY MATERIALS \\ [.2cm] ``An effective density matrix for vacua in asymptotically flat gravity'' } \\ [.2cm]
\vspace{0.05in}
{Temple He, Prahar Mitra, and Kathryn M. Zurek}
\end{center}

\setcounter{equation}{0}
\setcounter{figure}{0}
\setcounter{table}{0}
\setcounter{section}{0}
\setcounter{page}{1}
\makeatletter
\renewcommand{\theequation}{S\arabic{equation}}
\renewcommand{\thefigure}{S\arabic{figure}}
\renewcommand{\thetable}{S\arabic{table}}

\onecolumngrid

\section{The Path Integral Measure}

\noindent The goal of this appendix is to construct the measure that satisfies Eq.~\eqref{measure-constraint}, which we reproduce here for convenience:
\begin{align}\label{measure-app}
	\int [\CD\CC] e^{-S_\eff[\CC]} = 1.
\end{align}
First, we recall from Eqs.~\eqref{Lmax_identification} and \eqref{ortho-prod} that
\begin{align}\label{ortho-prod-app}
\begin{split}
S_\eff[\CC] &= \frac{\pi}{\a} \sum_{\ell=2}^{\ell_{\max}} \sum_{m=1}^{2\ell+1} (\ell-1)\ell(\ell+1)(\ell+2) \CC_{\ell m}^2 , \qquad \ell_{\max} = \sqrt{\frac{A}{\e^2}} - 1 .
\end{split}
\end{align}
Now, the path integral measure in general can be written as 
\begin{align}\label{ansatz}
\begin{split}
\int [\CD\CC] = \frac{1}{\text{vol}({\mathbb R}^4)} \exp \left( - a_1 - \frac{c}{6} \ln \frac{A}{\e^2} \right) \prod_{\ell=0}^{\ell_{\max}} \prod_{m=1}^{2\ell+1}  \frac{a_2}{(A/\e^2)^\l} \int_{-\infty}^\infty \dt \CC_{\ell m} , 
\end{split}
\end{align}
where $a_1,a_2,c,\l$ are all currently unspecified dimensionless parameters that we will determine to ensure the UV divergences are renormalized. This measure is manifestly invariant under supertranslations$/\mathbb R^4$ and the $SU(2)$ subgroup of $SL(2,\mathbb C)$ (it is an interesting question whether the form of this measure can be further fixed by imposing the full $SL(2,\mathbb C)$). Notice that Eq.~\eqref{ansatz} with Eq.~\eqref{measure-normalization} imply that the gauge-invariant Dirac delta function involving $\CC$, which excludes the $\ell=0,1$ gauge modes, is then in terms of spherical harmonic modes
\begin{align}
\begin{split}
\d(\CC-\CC') &=  \left[ \frac{(A/\e^2)^\l}{a_2}  \right]^4 \exp \left( a_1 + \frac{c}{6} \ln \frac{A}{\e^2}  \right) \prod_{\ell=2}^{\ell_{\max}} \prod_{m=1}^{2\ell+1} \d \left( \frac{a_2}{(A/\e^2)^\l} (\CC_{\ell m} - \CC'_{\ell m}) \right) .  
\end{split}
\end{align}
The constants $a_2$ and $\l$ in Eq.~\eqref{ansatz} renormalize $\CC$, whereas $a_1$ is associated to a counterterm action. Additionally, we divide by $\text{vol}({\mathbb R}^4)$ to account for the gauge symmetry given by Eq.~\eqref{gauge-freedom}.

As an aside, we note that $c$ parametrizes the logarithmic divergence in the measure that cannot be canceled by local counterterms. Indeed, $c$ is the central charge of the theory \cite{Gomis:2015yaa}. To see this, we note that we can think of the change in the radius $R$ of the sphere as a Weyl transformation given by
\begin{align}\label{weyl}
	\d_{\w} g_{ab} = 2 \o g_{ab} \quad\implies\quad \d_\w R = \w R .
\end{align}
Noting that our soft effective action is conformally invariant, only the nonlocal term will contribute to the trace anomaly. Using Eq.~(3.4.6) of \cite{Polchinski:1998rq}, we can compute the central charge $c_T$ of the theory to be
\begin{align}
\begin{split}
	\d_\w \bigg( - \frac{c}{6} \ln \frac{A}{\e^2} \bigg) = \frac{1}{2\pi} \int_{S^2} \dt^2x\, \sqrt{g} \w \la T^a{}_a \ra = - \frac{c_T}{24\pi} \int_{S^2} \dt^2x\,\sqrt{g} \w \CR_g =  - \frac{c_T}{3}  \w ,
\end{split}
\end{align}
where $\CR_g = 2/R^2$ is the Ricci scalar of the sphere, and we used the fact $\la T^a{}_a\ra = - c_T\CR_g/12$. Using $A=4\pi R^2$ and substituting in Eq.~\eqref{weyl}, we get $c_T = c$, proving our claim that $c$ is the central charge.

We now want to confirm that we can choose the parameters in Eq.~\eqref{ansatz} such that Eq.~\eqref{measure-app} holds. Substituting Eqs.~\eqref{ortho-prod-app} and \eqref{ansatz} into Eq.~\eqref{measure-app} and carrying out the resultant Gaussian integrals, we get
\begin{align}
\begin{split}
\exp \left( - a_1 - \frac{c}{6} \ln \frac{A}{\e^2} \right) \prod_{\ell=0}^{\ell_{\max}} \left[ \frac{ a_2}{(A/\e^2)^\l} \right]^{2\ell+1} \prod_{\ell=2}^{\ell_{\max}} \left[ \frac{\a }{ ( \ell - 1 ) \ell ( \ell + 1 ) ( \ell + 2 )  } \right]^{\frac{2\ell+1}{2}}  = 1 . 
\end{split}
\end{align}
Taking the logarithm of both sides, we get
\begin{align}
\begin{split}
\label{param-inter}
0 = - a_1 - \frac{c}{6} \ln \frac{A}{\e^2} + (\ell_{\max} + 1 )^2 \left[  \ln a_2 - \l \ln \frac{A}{\e^2} \right] + \frac{1}{2} \sum_{\ell=2}^{\ell_{\max}} (2\ell+1) \ln \frac{\a}{( \ell - 1 ) \ell ( \ell + 1 ) ( \ell + 2 )}  . 
\end{split}
\end{align}
We evaluate the sum in the last term in \texttt{Mathematica} and simplify using the identity (for $\text{Re}\,x>-1$)
\begin{align}\label{id}
\begin{split}
\p_s \zeta(s,x+1)\big|_{s=-1} &= \psi^{(-2)}(x) +  \frac{1}{2} x^2 + x\ln x - \frac{x}{2}\log(2\pi e ) + \frac{1}{12} - \ln \CA ,
\end{split}
\end{align}
where $\zeta(s,x)$ is the generalized Reimann zeta function, $\psi^{(\nu)}(x)$ the polygamma function, and $\CA \approx 1.2824$ is the Glaisher-Kinkelin constant. This implies
\begin{equation}
\begin{split}
\label{sum-result}
\frac{1}{2} \sum_{\ell=2}^{\ell_{\max}} (2\ell+1) \ln \frac{\a}{( \ell - 1 ) \ell ( \ell + 1 ) ( \ell + 2 )} &= \frac{1}{2} ( \ell_{\max} - 1 ) ( \ell_{\max} + 3 ) \ln \a - 4 \psi^{(-2)}(\ell_{\max}) \\
&\qquad  - \frac{1}{2} ( 2 \ell_{\max} + 1 ) \ln (\ell_{\max}+2) - \frac{3}{2} ( 2 \ell_{\max} - 1 ) \ln \ell_{\max}  \\
&\qquad  - 2 \ell_{\max} \big[\ell_{\max} + \ln (\ell_{\max}+1) - \ln (2 \pi e) \big]  + \frac{1}{2} \ln (432)  . 
\end{split}
\end{equation}
Substituting Eq.~\eqref{sum-result} into Eq.~\eqref{param-inter}, simplifying using Eq.~\eqref{ortho-prod-app}, and expanding about $A/\e^2 \gg 1$, we find the condition
\begin{align}
\begin{split}
0 &= \frac{A}{\e^2} \left[ - ( \l + 1 ) \ln \frac{A}{\e^2} + \ln ( a_2 \sqrt{\a} e ) \right] + \frac{8-c}{6} \ln \frac{A}{\e^2} - a_1 - 2 \ln \a - 4 \ln\CA  - 2  + \frac{1}{2} \ln 432  + \CO(\e^2/A)  .
\end{split}
\end{align}
It is then straightforward to check that this equality is satisfied given
\begin{equation}
\begin{split}\label{measure-constants}
\l = -1 ,  \qquad a_1 = - 2 + \frac{1}{2} \ln 432 - 4 \ln \CA - 2 \ln \a  , \qquad a_2 = \frac{1}{\sqrt{\a} e }  , \qquad c = 8 . 
\end{split}
\end{equation}
This completes our verification that a measure satisfying Eq.~\eqref{measure-app} exists.

\newpage

\section{Thermal Partition Functions}

We review the usual calculation of the thermal partition function for the free particle and the simple harmonic oscillator (SHO). For power counting reasons, we keep all factors of $\hbar$ explicit.

\subsection{Free Particle on a Line} 
\noindent The Hamiltonian of a free particle on a line with mass $m$ is given by
\begin{equation}
\begin{split}
{\wh H}_{\text{line}} = \frac{{\wh P}^2}{2m} , \qquad [ {\hat X} , {\hat P} ] = i \hbar , \qquad {\hat X} \in {\mathbb R}.
\end{split}
\end{equation}
Momentum (and energy) eigenstates with eigenvalue $p \in {\mathbb R}$ ($E_p = \frac{p^2}{2m}$) are denoted by $|p\ra$. These are $\d$-function normalized, in that $\braket{p}{p'} = 2\pi \hbar \d(p-p')$. The thermal partition function is given by
\begin{equation}\label{Z-line}
\begin{split}
	Z_\text{line}(\b) = \Tr \big[ e^{- \b {\hat H}_\text{line}} \big] = \la p | p \ra \int_{-\infty}^\infty \frac{\dt p}{2\pi \hbar} e^{-  \frac{\b p^2}{2m}  }  = \sqrt{ \frac{m}{2\pi \b \hbar^2} } \la p | p \ra .
\end{split}
\end{equation}
The mean and variance of any modular Hamiltonian $\hat K(\b)$ is given in terms of the partition function $Z(\b)$ by the standard thermodynamic identities
\begin{align}\label{thermal-id}
\begin{split}
	\la {\hat K}(\b) \ra = ( 1 - \b \p_\b ) \ln Z(\b)  , \qquad \la \D {\wh K}^2(\b) \ra = \b^2\p_\b^2 \ln Z(\b)  .
\end{split}
\end{align}
Substituting Eq.~\eqref{Z-line} into the above identities, we obtain
\begin{equation}
\begin{split}
\label{KT-line_app}
\la {\hat K}_\text{line}(\b) \ra  =  \frac{1}{2} + \ln \bigg( \sqrt{ \frac{m}{2\pi\b \hbar^2}} \la p | p \ra \bigg) , \qquad \la \D {\wh K}_\text{line}^2(\b) \ra = \frac{1}{2} .
\end{split}
\end{equation}
Since $|p\ra$ is non-normalizable, we see the mean has a logarithmic divergence.

\subsection{Free Particle on a Ring} 
\noindent The logarithmic divergence in Eq.~\eqref{KT-line_app} can be regulated by making ${\hat X}$ compact. To see this, consider the Hamiltonian for a free particle on a ring of radius $R$ with mass $m$, so that
\begin{equation}
\begin{split}
{\wh H}_{\text{ring}} = \frac{{\wh P}^2}{2m} , \qquad [ {\hat X} , {\hat P} ] = i \hbar , \qquad {\hat X} \sim {\hat X} + 2\pi R .
\end{split}
\end{equation}
Momentum (and energy) eigenstates with eigenvalue $p_n=\frac{n \hbar}{R}$ (and $E_n = \frac{n^2 \hbar^2}{2m R^2}$), where $n \in {\mathbb Z}$, are denoted by $\ket{n}$. These are normalized as $\la n | n' \ra = \d_{nn'}$. The thermal partition function is now
\begin{equation}
\begin{split}
Z_\text{ring}(\b) = \Tr \big[ e^{- \b {\hat H}_\text{ring}} \big] = \sum_{n=-\infty}^\infty e^{- \frac{\b \hbar^2}{2m r^2} n^2  } = \vartheta _3\left( 0 , e^{-\frac{\b \hbar^2}{2m R^2} }\right) . 
\end{split}
\end{equation}
In this case, we find 
\begin{equation}
\begin{split}
\label{KT_ring_app}
\la {\hat K}_\text{ring}(\b) \ra &= ( 1 - \b \p_\b ) \ln Z_\text{ring}(\b) = - \b \p_\b \ln \vartheta _3\Big( 0 , e^{-\frac{\b \hbar^2}{2m R^2} }\Big) + \ln \vartheta _3\Big( 0 , e^{-\frac{\b \hbar^2}{2m R^2} }\Big) . 
\end{split}
\end{equation}
We see that the logarithmic divergence in Eq.~\eqref{KT-line_app} is regulated by the Jacobi theta function. In the limit $R \to \infty$ where the ring approximates the line, we recover the logarithmic divergence since
\begin{equation}
\begin{split}
	\ln \vartheta _3\Big( 0 , e^{-\frac{\b \hbar^2}{2m R^2} }\Big) \quad  \xrightarrow{R \to \infty} \quad \ln \left( \sqrt{ \frac{ m }{ 2\pi \b \hbar^2 } } ( 2 \pi R ) \right) .
\end{split}
\end{equation}
Comparing this to Eq.~\eqref{KT-line_app}, we see that putting the free particle on a ring regulates $\la p | p \ra$ to $2\pi R$.

\subsection{Simple Harmonic Oscillator}

Another way to regulate the logarithmic divergence in Eq.~\eqref{KT-line_app} is to add a potential term to the Hamiltonian. To see this, consider the simple harmonic oscillator (SHO) with mass $m$ and frequency $\w$ is given by
\begin{align}\label{SHO-ham}
	{\wh H}_{\sho}= \frac{{\wh P}^2}{2m} + \frac{1}{2} m \w^2 {\wh X}^2, \qquad [{\wh X} , {\wh P} ] = i\hbar , \qquad {\hat X} \in {\mathbb R}.
\end{align}
In this case, even though ${\hat X}$ is non-compact, large eigenvalues of ${\hat X}$, namely when $|x| \gg \sqrt{\hbar/(m\o)}$, are highly suppressed due to the quadratic potential. This effectively makes ${\hat X}$ compact, which then regulates the logarithmic divergence. 

Using the fact the energy spectrum of the SHO is given by $E_n = (n + \frac{1}{2})\hbar\w$ where $n \in {\mathbb Z}_{\geq 0}$, the thermal partition function for the SHO is given by
\begin{align}\label{Z-beta}
\begin{split}
Z_\sho(\b) &\equiv \Tr\big[ e^{-\b {\wh H}_\sho}\big] = \sum_{n=0}^\infty \la n | e^{-\b {\wh H}} | n \ra = \sum_{n=0}^\infty  e^{\b\hbar\w(n + \frac{1}{2})} = \frac{e^{\frac{\b\hbar\w}{2}}}{e^{\b \hbar \w}-1} .
\end{split}
\end{align}
Having computed the partition function $Z_\sho(\b)$, we can evaluate the mean and variance of the modular Hamiltonian using 
\begin{align}
\begin{split}
\label{K-beta-fin}
\la {\wh K}_\sho(\b) \ra &= (1-\b\p_\b) \ln Z_\sho(\b)  = \frac{\b \hbar \o}{e^{\b \hbar \o}-1} - \ln ( 1 - e^{- \b \hbar \o} )  ,  \\
\la \D {\wh K}_\sho^2(\b) \ra &= \b^2\p_\b^2 \ln Z_\sho(\b) =  \frac{(\b \hbar \o)^2 e^{\b \hbar \o}}{(e^{\b \hbar \o}-1)^2} . 
\end{split}
\end{align}
Notice that in the main text, the mean and variance of $ {\wh K}_s$ we computed is analogous to us taking $\w \to 0$ in Eq.~\eqref{K-beta-fin}, as our $\CC,\CN$ modes are at exactly zero energy and behave as free scalar fields. In that case, we get 
\begin{align}
\begin{split}
\lim_{\w \to 0} \la {\wh K}_\sho(\b) \ra &= 1 - \ln(\b\hbar\w) , \qquad \lim_{\w \to 0} \la \D {\wh K}_\sho^2(\b) \ra  = 1.
\end{split}
\end{align}
This is precisely the analog of Eqs.~\eqref{eq:Ks} and \eqref{final}.

Thus far, we evaluated the mean and variance of the modular Hamiltonian of a single SHO in a thermal background. A typical field theory consists of a continuum of SHOs. For a non-interacting continuum of SHOs with frequency density $n(\o)$, the mean and variance of the modular Hamiltonian are given by
\begin{equation}
\begin{split}
\label{barK-12pts2}
\la {\wh K}_\sho(\b) \ra &= \int_0^\infty \dt\w\, n(\w) \bigg[ \frac{\b \hbar \o}{e^{\b \hbar \o}-1} - \ln ( 1 - e^{- \b \hbar \o} ) \bigg]  ,\qquad \la \D {\wh K}_\sho^2(\b) \ra = \int_0^\infty \dt\w\, n(\w)\frac{(\b \hbar \o)^2 e^{\b \hbar \o}}{(e^{\b \hbar \o}-1)^2}  .
\end{split}
\end{equation}
This follows from the fact that for non-interacting systems, the mean and variance of the total modular Hamiltonian is the sum of the mean and variance of the modular Hamiltonian of each system, respectively. For a free massive field theory in $D$ dimensions, we have $n(\o) = n_0 \hbar \o ( \hbar^2 \o^2 - m^2 )^{\frac{1}{2}(D-3)} \t(\hbar  \o-m)$, where $\th$ is the Heaviside step function. In particular, notice that for massless fields in two dimensions, $n(\o)=n_0$ is independent of $\o$, and so Eq.~\eqref{barK-12pts2} reduces to
\begin{equation}
\begin{split}
\la {\wh K}_\sho(\b) \ra &=  \la \D {\wh K}_\sho^2(\b) \ra = \frac{\pi^2 n_0}{3 \b \hbar} .
\end{split}
\end{equation}

\end{document}